%% file: ms.tex
\documentclass[apj]{emulateapj}
\usepackage{apjfonts}
\usepackage{wrapfig}
\usepackage{amsmath}

\shorttitle{DETECTION OF CHROMATIC MICROLENSING IN Q~2237+0305 A}
\shortauthors{MOSQUERA, MU\~{N}OZ $\&$ MEDIAVILLA}

\begin{document}
\slugcomment{Accepted for publication in The Astrophysical Journal}

\title{DETECTION OF CHROMATIC MICROLENSING IN Q~2237+0305 A}


\author{A.M. Mosquera\altaffilmark{1}, J.A. Mu\~noz\altaffilmark{1}, and E. Mediavilla\altaffilmark{2}}

\altaffiltext{1}{Departamento de Astronom\'{\i}a y Astrof\'{\i}sica, Universidad
       de Valencia, E-46100 Burjassot, Valencia, Spain}
\altaffiltext{2}{Instituto de Astrof\'{\i}sica de Canarias, E-38200 La Laguna,
       Tenerife, Spain}


\begin{abstract}
We present narrowband images of the gravitational lens system Q~2237+0305
made with the Nordic Optical Telescope in eight different filters
covering the wavelength interval  3510-8130 \AA. Using  point-spread function
photometry fitting we have derived the difference in magnitude versus wavelength 
between the four images of Q~2237+0305. At $\lambda=4110$ \AA, the wavelength range 
covered by the Str\"omgren-v filter coincides with the position and width of the CIV
emission line. This allows us to determine the existence of microlensing in the 
continuum and
not in the emission lines for two images of the quasar. Moreover, the brightness of 
image A shows a significant variation with wavelength which can only be 
explained as consequence of chromatic microlensing. To perform a 
complete analysis of this chromatic event our observations were used together
with Optical Gravitational Lensing Experiment light curves. Both data sets 
cannot be reproduced by the simple phenomenology 
described under the caustic crossing approximation; using more realistic 
representations of microlensing at high optical 
depth, we found solutions consistent  with simple thin disk models 
($r_{s}\varpropto \lambda^{4/3}$); however, other accretion disk 
size-wavelength relationships also lead to good solutions. 
New chromatic events from the ongoing narrow  band photometric monitoring 
of Q~2237+0305 are needed to accurately constrain the physical properties 
of the accretion disk for this system.

\end{abstract}


\section{Introduction}

Gravitational lensing is independent of wavelength (Schneider, Ehlers \& Falco 1992). However, in many 
gravitationally lensed quasars, differences in color between the images are observed. 
These chromatic variations  could be produced by two effects: 
differential extinction in the lens galaxy and chromatic microlensing.
When each image's light crosses the interstellar medium of the lens galaxy it may be affected in different amounts by patchily 
distributed dust. This results in differential extinction between  pairs of images, and makes 
possible the determination of the extinction law of the lens galaxy (Nadeau et al. 1991; Falco et al. 1999; 
Motta et al. 2002; Mu\~noz et al. 2004; Mediavilla et al. 2005; 
El\'{i}asd\'{o}ttir et al. 2006). The other chromatic phenomenon arises when stars or compact objects in the lens galaxy 
are nearly aligned with the line of sight between the quasar image and the observer. Due to the relative motion between the 
quasar, the lens and the observer, the quasar image undergoes a magnification or
demagnification known as microlensing (Schneider, Kochanek $\&$  Wambsganss 2006 and references therein). 
Therefore, fluctuations in the brightness of a quasar image will be a combination of 
the intrinsic quasar variability and the microlensing 
from the stars or compact objects in the lens galaxy. 
The intrinsic variability of the lensed quasar will appear in all images 
with certain time delay due to the different light travel times. Once this delay is determined, the light curves of the 
different images can be shifted, and then subtracted. The remaining fluctuations can be assumed to be 
caused only by  microlensing. The microlensing 
magnification depends on the angular size of the source, in this case on the accretion disk of the quasar. Because the accretion disk is
hotter closer to the black hole, and because the emission of the accretion disk depends on temperature, different magnifications
may be observed at different wavelengths. 
This effect is known as chromatic microlensing, and it offers unprecedented perspectives into the 
physical properties of accretion disks (Wambsganss $\&$ Paczynski 1991). Its detection in a lens system will lead to accurate constraints
in the size-wavelength scaling. 

Different authors have made some attempts to detect chromatic microlensing and use it with different scopes. 
However, in many cases the detection of chromatic microlensing was rather ambiguous, since 
in general it is not easy
to disentangle this effect from others with similar observational signatures 
(Wisotzki et al. 1993; Nadeau et 
al. 1999; Wucknitz et al. 2003; Nakos et al. 2005). The first significant applications 
of chromatic microlensing
appeared in a very recent work of Poindexter et al. (2008). They  have used chromatic microlensing in 
order to determine the size of the accretion disk of HE 1104-1805 as
well as its size-wavelength scaling. More recently, Anguita et al. (2008) have 
also made these kinds of studies 
in Q~2237+0305. In both cases they found solutions consistent with the simple 
thin disk model (Shakura $\&$ Sunyaev 1973), 
but stronger constraints would be needed to accurately determine the size of the accretion disk.

In this work, we present a chromatic microlensing detection in one of the 
images of Q~2237+0305 (Huchra et al. 1985). 
This lens system has very good properties to study microlensing events. 
The light coming from a distant quasar 
at $z_S=1.695$ is deflected by a nearby spiral galaxy 
($z_L=0.039$) forming four quasar images nearly symmetrically 
distributed around the lens center. Since the light passes through 
the central part of the galaxy, where the microlensing optical depth is very large, 
high magnification events (HMEs) occur very 
frequently. Moreover, since the galaxy has a very low redshift, and 
due to the symmetry in the positions of the images,
time delays are expected to be very small ($< 1$ day; Vakulik et al. 2006; Koptelova et al. 2006). 
Therefore, time delay corrections are not needed to study microlensing in this 
lens system, and the flux ratio between 
the images will remove the contribution of the intrinsic quasar variability in the light curves.

Specifically, we analyze a particular chromatic microlensing event, and study the physical 
scenarios in which this effect can 
be produced. The results obtained in this work are the first step toward obtaining precise
constraints in the physical properties of the quasar accretion disk.
In \S 2 the observations and the data analysis are presented. The data fitting techniques and different
models to describe the observed chromatic microlensing are discussed in \S 3. A summary of 
the main conclusions
appears in \S 4.

\section{Observations and Data analysis}

On the nights of August 26 and 28 2003, we observed Q~2237+0305 with the 2.56 m Nordic Optical Telescope (NOT) located at 
the Roque de los Muchachos Observatory, La Palma (Spain), using the 2048$\times$2048 ALSFOC detector. Its spatial scale is 
0.188 arsec/pixel. Seven narrow filters plus the wide I-Bessel were used. The whole set covered the wavelength interval 
3510-8130 \AA. Table 1 shows a log of our observations.

\input{tab1.tex}

Among the filters that we used only two were affected by the emission lines of the quasar.  
The Str\"omgren-u filter 
was affected by almost 40\% by the Ly$\alpha$ emission line, where the emission lines of the 
quasar and its nearby continuum were modeled according to the SDSS quasar composite spectrum 
(Vanden Berk et al. 2001) 
to perform this estimation. At $\lambda=4110$\AA \ the wavelength range covered by 
the Str\"omgren-v filter coincides with 
the position and width of the CIV emission line.

The data were reduced using standard procedures with IRAF packages, and PSF 
photometry fitting  was used to derive the difference in magnitude versus wavelength 
between the four images of Q~2237+0305. The galaxy bulge was 
modeled with a de Vaucouleurs profile, and the quasar images as point sources. 
This model was convolved with different PSFs observed simultaneously with the lens 
system in each of the frames, and compared to the 
image through $\chi^2$ statistics ( McLeod et al. 1998; Leh\'{a}r et al. 2000).  
Due to the good seeing conditions (0''.6 in I band), the results of the photometry 
were excellent even in the bluest filters.

\begin{figure} 
\centerline{\includegraphics[width=3.5in]{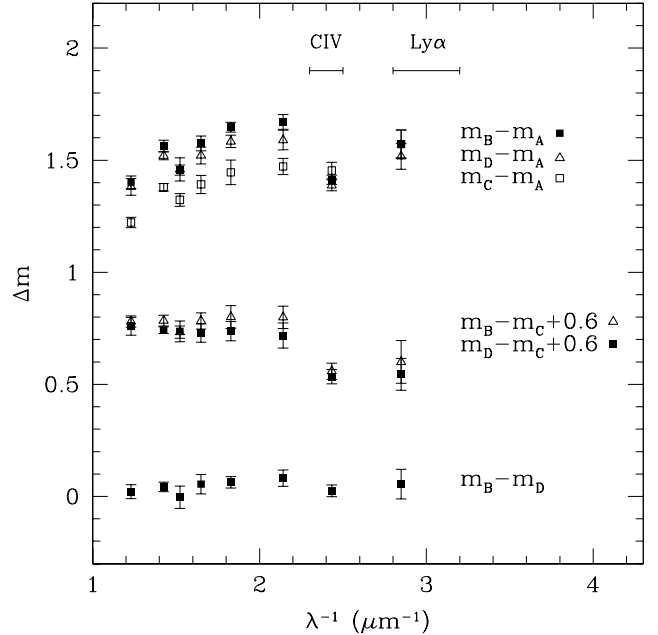}}
\caption{\label{fig:curvas2} Magnitude differences as a function of the observed wavelength 
for Q~2237+0305 (NOT data). Images A and C show a clear signature of
microlensing in the continuum but not in the emission lines. Moreover, 
the chromatic variation observed in image A could only be explained 
as a consequence of chromatic microlensing. The wavelength widths covered by the 
emission lines that affect two of our filters (CIV and Ly$\alpha$) are indicated.}
\end{figure}

Table 2 presents the measurements of the differences in magnitude in each filter. 
In Figure  \ref{fig:curvas2}, we have plotted six magnitude  differences versus 
wavelength (three of which are independent) 
between the four images of the quasar.

\input{tab2.tex}

\begin{figure*}
\centerline{\includegraphics[height=5.0in]{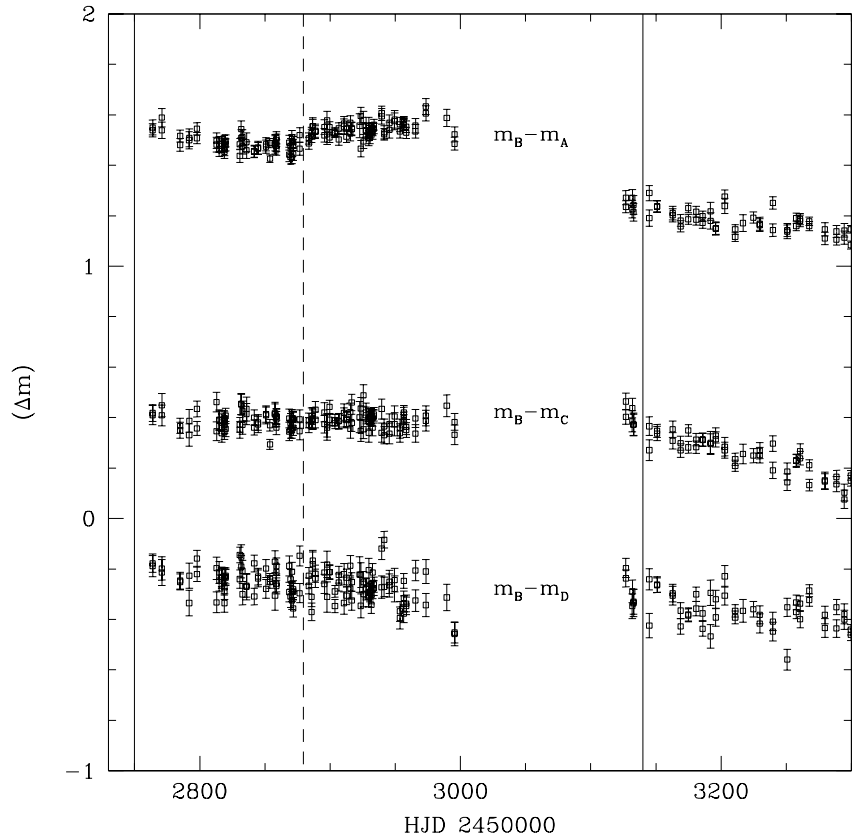}}
\caption{\label{fig:ogle_Bdiffs} Magnitude differences between Q~2237+0305 images obtained from OGLE V-band light curves.
The dashed line indicates our observation date, and the solid lines limit the selected time interval in which the study was performed.}
\end{figure*}

The difference in magnitude between B and D does not show any chromatic variation, and it is reasonable to 
suppose that neither B nor D is undergoing flux  variations with wavelength. The differences 
$m_B-m_C$, and $m_D-m_C$, show that both curves have the same trend, and this implies that the magnitude
of image C varies with wavelength (consistent with our hypothesis that neither 
image B nor D suffers of chromaticity). In these curves it is also observed that there is no 
magnitude variation 
with the wavelength in the region between 4670 \AA \ and  8130 \AA, but at the wavelengths 
corresponding to 
the Str\"omgren-v and Str\"omgren-u filters, which are the only ones affected by emission lines, 
we observed a deviation 
from the trend. As we stated above the Str\"omgren-u filter was affected by almost 40\% by  
the Ly$\alpha$ emission line, 
and at $\lambda=4110$ \AA, the wavelength range covered by the Str\"omgren-v filter coincides 
with the position 
and width of the CIV emission line. Then, the dependence of  $m_B-m_C$ or $m_D-m_C$ with 
wavelength observed under 
our filter set configuration is the observational signature expected for image C being 
affected by microlensing in the 
continuum but not in the emission lines. The fact that microlensing is absent in the 
emission lines confirms the 
relatively large size of the broad-line region (BLR) in Q~2237+0305 according to the 
existing relation between the size of the 
BLR and the intrinsic luminosity of the quasar (Kaspi et al. 2000, Abajas et al. 2002). 

Finally, we analyzed the differences between image A and the other ones. Again, as the trend of 
the curves is the same 
for each difference, we conclude that A is the image which undergoes flux variations. 
As the points corresponding to 
Str\"omgren-v and Str\"omgren-u filters deviate from the respective trend, we can 
conclude again that image A is undergoing 
microlensing in the continuum but not in the emission lines. An even more interesting 
result appears from analyzing 
the points corresponding to the continuum, where image A shows a significant variation 
with wavelength. In principle, this 
effect could be produced by extinction and/or chromatic microlensing. But as image 
A is the one undergoing the flux 
variation, if it were due to extinction the slope of the curve should be negative, 
and not positive as it is observed in the plot. 
Therefore, we have strong evidence of chromatic microlensing in image A. 

Since we have observations 
of only two consecutive nights in that period, to know what was the 
signature of the microlensing in the system, we used 
the public Optical Gravitational Lensing Experiment(OGLE)
\footnote{\label{oglelink} http://bulge.princeton.edu/$\sim$ogle/} data 
(Wo\'zniak et al. 2000). Since 1999 January 
OGLE has been monitoring Q~2237+0305 in the V-band. In Figure  \ref{fig:ogle_Bdiffs}, 
three independent magnitude differences 
between the quasar images obtained from their data are plotted. In the time 
interval between HJD 2452764 and HJD 2453145 (indicated 
in Figure  \ref{fig:ogle_Bdiffs} between solid lines) the differences ($m_B-m_C$) 
and ($m_B-m_D$) are almost constant. 
This would imply that the flux variations of the images B, C, and D are caused mainly by intrinsic 
variability of the quasar, which agrees with our previous hypothesis that neither B, C nor
D is undergoing chromaticity originated by an HME. The fact that in our observations 
we have detected (nonchromatic)
microlensing in image C is also in agreement with OGLE data, since a constant 
micro-magnification value during this
period justifies both observations. Therefore, the brightness fluctuations 
($\sim 0.3$ mag) observed in the difference ($m_B-m_A$) 
can be assumed to be due only to microlensing-induced variability in image A. 
In other words, from Figure  \ref{fig:ogle_Bdiffs} we
can say that image A was in fact undergoing an HME on the dates of our observations which 
correspond to HJD  2452878.5 and 2452880.5.

Summarizing we can say that, based on the 2003 NOT observations, images C and A are undergoing
microlensing in the continuum but not in the emission lines, and that 
image A is also affected by chromatic microlensing. In the next section this unambiguous detection of 
chromatic microlensing is used, along with the OGLE data from HJD 2452764 to 
HJD 2453145, to explore the phenomenology 
of the detected microlensing event and to study the physical properties of the accretion disk.

\section {Data fitting}

\subsection{Chromatic microlensing and caustic crossing approximation}

\begin{figure} 
\centerline{\includegraphics[width=3.5in]{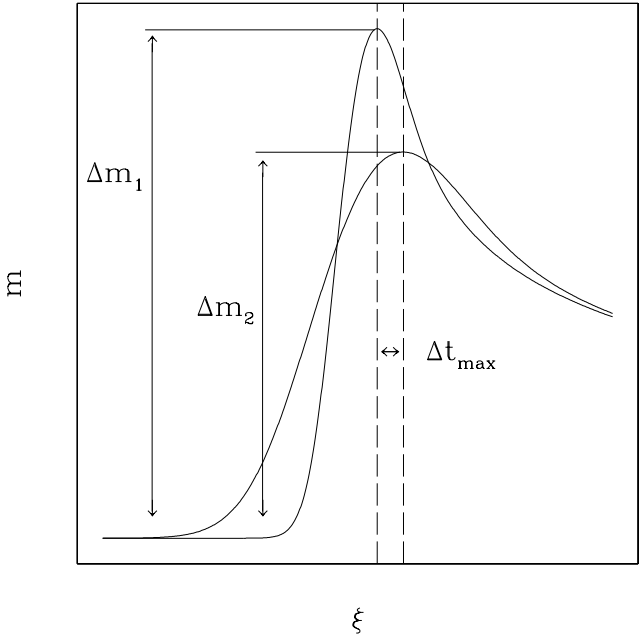}}
\caption{\label{fig:mic_chrom} Typical observable effects of chromatic microlensing. The amplitude
of the microlensing is different at different wavelengths due to the expected dependence of the 
accretion disk size with wavelength. In addition, the maximum magnification is reached at a different
instant with a time delay $\Delta t_{max}$ (see the text).}
\end{figure}

To review some  basic aspects of chromatic microlensing we will start recalling the 
simplest model for microlensing, the caustic crossing approximation. Taking a Cartesian coordinate
frame in which the caustic lies along the $y$-axis, the magnification  
at a distance $x$ is proportional to the well-known expression $ 1+\beta\ H(x)/\sqrt{x}$ \ 
(Schneider $\&$ Weiss 1987). 
Here $H$ represents the Heaviside step function, and $\beta$ measures the caustic strength. 
Assume a brightness 
profile for the quasar accretion disk given by
\begin{equation}\label{gauss_I}
I(r)=\frac {I_{0}}{2 \pi r_s^2} \Psi(r/r_s),
\end{equation}
where $r_{s}$ represents a typical size scale of the quasar accretion disk; 
provided a global factor (or a global constant in magnitudes) containing the magnification
background, the magnification of the 
quasar accretion disk is
\begin{equation}\label{amplification}
\mu \sim 1+\frac{\beta} {\sqrt{r_{s}}} \ \phi(\xi),
\end{equation}
where $\xi \equiv \frac{x_{0}}{r_{s}}$ is the distance between the center of the disk and 
the caustic, $x_{0}$, in units of
the scale radius $r_s$, and it can take negative or positive values to distinguish both 
sides of the caustic. 
Analytical expressions
can be straightforwardly obtained for several accretion disk models, {\it e.g.}, 
Gaussian
\footnote{$\Psi(r/r_s)=e^{-\frac{r^2}{2r^2_s}}\rightarrow \phi(\xi)= 2^{-3/2} \,
\pi ^{1/2} \, e^{-\xi^2/4} \,|\xi|^{1/2}
\left[I_{-\frac{1}{4}}(\frac{\xi^2}{4}) \, +\, \frac{\xi}{|\xi|} \,  
I_{\frac{1}{4}}(\frac{\xi^2}{4} )\right]$ where $I$
is the modified Bessel function of the first kind.}
 (Schneider $\&$ Weiss 1987) or power 
laws (Shalyapin 2001). If a constant 
relative source plane velocity is considered between the disk and the caustic, 
then $x_{0}=v \ (t-t_{0})$,  
where $t_{0}$ is the instant of time at which the center of the disk is located at the caustic 
position. Therefore we can see that an HME due to a single caustic 
crossing is, in addition to the crossing time $t_{0}$, a function of only two parameters:
$\beta'\equiv \beta/\sqrt{r_{s}}$ and $r'_s \equiv r_{s}/v$. Due to
this degeneracy, the disk size $r_{s}$ and the velocity $v$ cannot be determined independently but
only the ratio $r_{s}/v$.

Figure  \ref{fig:mic_chrom} shows the typical observable effects of chromatic microlensing. 
Because the size of the accretion disk is expected to change with wavelength, the amplitudes of the 
HME events observed at different wavelengths are different. The maximum magnification, in magnitudes, relative to the
local average microlensing magnification background (see Figure  \ref{fig:mic_chrom}) will be
\begin{equation}\label{deltam0}
\Delta m_{max}=-2.5\ \mathrm{Log}  \left[1+\frac{\beta}{\sqrt{r_{s}}}\ \phi(\xi_{max})\right].
\end{equation}
The value of $ \phi(\xi_{max})$ depends on the brightness profile model. For instance, 
for a Gaussian model $ \phi(\xi_{max}\simeq 0.765)\simeq 1.021$. Hence, given a brightness profile model, the observable
$\Delta m_{max}$ provides a direct estimation of the dimensionless parameter $\beta/\sqrt{r_s}$.

If an HME is monitored  at two different wavelengths, then using equation (\ref{deltam0}) the ratio
between the different corresponding sizes is
\begin{equation}\label{sratio}
\frac{r^1_{s}}{r^2_{s}}= \left[\frac{10 ^{-\Delta m^2_{max}/2.5}-1}{10 ^{-\Delta m^1_{max}/2.5}-1}\right]^{2}
\end{equation}
and it is very remarkable that, provided that equation (\ref{amplification}) is satisfied, this expression
is independent of the brightness profile of the accretion disk. As an example we can use the
recent results obtained by Anguita et al. (2008), where they analyzed another HME observed in the 
1999 OGLE campaign. From Figure 2 in their 
paper it is possible to estimate that $\Delta m^g_{max} \sim -0.8$ (see also OGLE $^{\ref{oglelink}}$
data for a better sampling) 
and $\Delta m^r_{max} \sim -0.57$. Therefore, applying equation (\ref{sratio})
{\bf $r^r_s/r^g_s \sim2.5$}. This value is slightly different from the ratio obtained in their paper where they use 
a magnification pattern instead of a caustic crossing.

\begin{figure} 
\centerline{\includegraphics[width=3.5in]{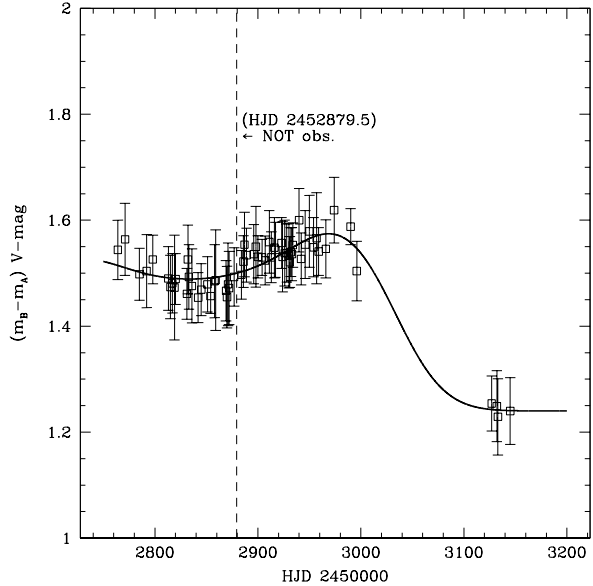}}
\caption{\label{fig:ogle} OGLE V-band ($m_B-m_A$) data. The best fit using
a simple double-caustic crossing approximation and a Gaussian brightness profile is shown. The dashed line
indicates our observation date.}
\end{figure}

Another interesting property is obtained with the observable $\Delta t_{max}$, {\it i.e.}, 
the difference in time to
reach the maximum magnification for each source size (see Figure \ref{fig:mic_chrom}). 
It is straightforward to obtain
\begin{equation}
\Delta t_{max}=\xi_{max} \  r'^{1}_{s} \left[1-\frac{r^2_{s}}{r^1_{s}}\right].
\end{equation}
Although it depends on the quasar disk model, it allows us to determine directly the 
size-velocity ratio. Unfortunately the light curve shown in Figure  2 of Anguita et al. (2008) has not 
the temporal sampling to accurately estimate $\Delta t_{max}$, but assuming a value $\sim 10$ days we would find 
the order of magnitude for $r'^g_s \equiv r^g_{s}/v \sim 20$ day$^{-1}$.

\subsection{Double-caustic crossing approximation}

We start using the simple model of the caustic crossing approximation with the aim of 
finding a simultaneous fit of
the difference $(m_B-m_A)$ obtained from the OGLE V-band light curves 
(over the temporal range from HJD 2452764 to HJD 2453145) (Figure  \ref{fig:ogle}) and our 
chromatic observation. As explained before, in this case, the fluctuations in 
 $(m_B-m_A)$ correspond only to microlensing-induced variability in image A. Because of the 
shape of the data we assumed that the image A of the quasar is undergoing a double-caustic crossing event, 
and we considered a Gaussian brightness profile\footnote{ Other 
brightness profiles were also used, 
as the thin accretion-disk approximation ({\it e.g.} Kochanek 2004), or power laws ({\it e.g.} 
Shalyapin et al. 2002), but the differences in the results produced by these other 
profiles are not significant.}
 for
the accretion disk $I(r)=\frac{I_{0}}{2\pi r_{s}^{2}} \exp \left[-r^{2}/2r^{2}_{s}\right]$. 
The main parameters involved in our fitting are: 
the ratio between $r_{s}$ and the relative caustic-disk 
source plane velocity, $r_{s}'=r_{s} / v$,  
the dimensionless caustic stretch divided by $\sqrt{r_{s}}$, $\beta_{i=1,2}'$, the time 
corresponding to each caustic crossing, $t_{i=1,2}$, 
and the relative microlensing magnification background between each side of the caustics, 
where the subindex $i=1,2$ refers to 
each of the caustics.

\begin{figure} 
\centerline{\includegraphics[width=3.5in]{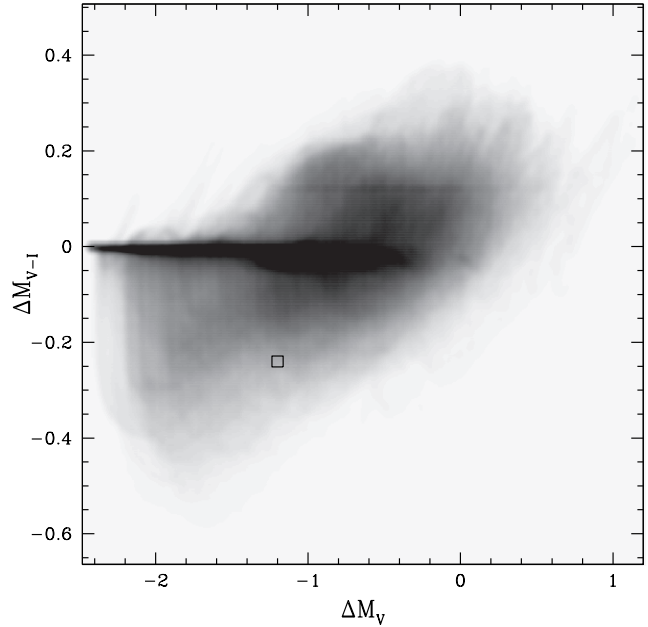}}
\caption{\label{fig:idl} Image A ``chromatic map''. $\Delta M_{V}$ 
represents the microlensing magnification in the V-band, 
and $\Delta M_{V-I}$ is the difference in magnitude $m_{V}-m_{I}$ 
due to chromatic microlensing. The open square corresponds to 
the observed values. The different grayscales in the map 
correspond to different values of the probability of occurrence.}
\end{figure}

The estimation of the model parameters was carried out with a $\chi^2$ minimization 
fitting method. Due to the scatter of the OGLE data, the points were binned to have only 
one point per day, which value corresponds to the averaged binned points, and which associated error corresponds
to the dispersion calculated in each day. Because the first caustic crossing event is not completely sampled, the 
parameters associated with it cannot be determined; however the ones associated with the second caustic can be bounded 
($\beta_2=0.3\pm0.1 $, $t_2=3008\pm12$). Figure  \ref{fig:ogle} shows the best fit that corresponds to $\chi^2/N_{f}=0.22$, 
where $N_{f}$ is the number of degrees of freedom. 

Although this simple model produces a good light curve fitting for the OGLE data, it cannot
reproduce the observed chromatic effect. Taking as reference the filters 
Str\"{o}mgren-y\footnote{The central wavelengths of the Str-y and V-band filters almost 
coincide. Therefore hereafter we will 
consider Str-y  and V-band  filters as equivalents to simplify the comparisons between our data set 
and that obtained by OGLE.} and I-band, 
we observe a chromaticity 
$(m^V_{A}-m^I_{A})= -0.24\pm 0.05$ mag . However, assuming a radius-wavelength scaling that goes as $r_{s}\varpropto \lambda^{4/3}$ (thin-disk 
model), the chromaticity that can be produced by this double-caustic approximation model on 
the day of our observation (HJD 2452879.5) 
is $\sim -0.01$ mag, and a maximum chromaticity of $\sim -0.25$ mag is only achieved for an 
unrealistic ratio $r^I_{s}/r^V_{s} \gtrsim 10^2$.
The limitation of the caustic crossing approximation in reproducing
the observed chromatic microlensing magnification is due to the fact that the induced chromaticity is
only a function of the caustic stretch $\beta/\sqrt{r_s}$, and it is
independent of the microamplification background. However, as we will see in the next subsection, 
this is not the case in a more realistic scenario.

This is a good example which confirms, as stated by Kochanek (2004), that 
due to the high microlensing optical 
depth of Q~2237+0305 it is unlikely that the system undergoes a ``clean'' 
caustic crossing. In this sense the model
that we used here is too simple to approach the real behavior of the system. 
The appropriate way to find a solution is to 
consider a more realistic physical scenario, {\it i.e.}, a magnification pattern is
needed to describe the observational signatures in this case.

\subsection {Modeling microlensing from magnification maps}

Magnification maps for images A and B would be needed to find trajectories that reproduce
($m_B-m_A$) V-band OGLE difference during the period of interest. As stated before, in the time interval
between HJD 2452764 and HJD 2453145, the brightness fluctuations observed in $(m_B-m_A)$ are due only 
to microlensing in image A, since image B is mainly affected by intrinsic variability of the quasar.
Therefore, the behavior of image B can be represented with a constant magnification value, 
and magnification patterns for this image are not needed. Its contribution to the $(m_B-m_A)$ magnitude
difference, as well as any extinction correction or uncertainties in the macromodels will be included in
a constant parameter, $m_0$, that will shift image A microlensing fluctuations.  

We have built magnification patterns for image A using the inverse ray shooting technique 
(Wambsganss 1990, 1999) and the inverse polygon mapping (Mediavilla et al. 2006). It 
was assumed that all the mass 
responsible for microlensing is in compact objects of 1 $M_{\odot}$, and the values for the 
convergence of $\kappa=0.36$ and for the 
shear of $\gamma=0.4$ (Schmidt et al. 1998) were 
considered. The map dimensions are of $4096 \times 4096$ pixels which correspond to physical dimensions 
of $20 \ r_{E} \times 20  \ r_{E}$. Therefore the magnification pattern has a resolution of $0.005$ $r_{E}/$pixel.

\begin{figure} 
\centerline{\includegraphics[width=3.5in]{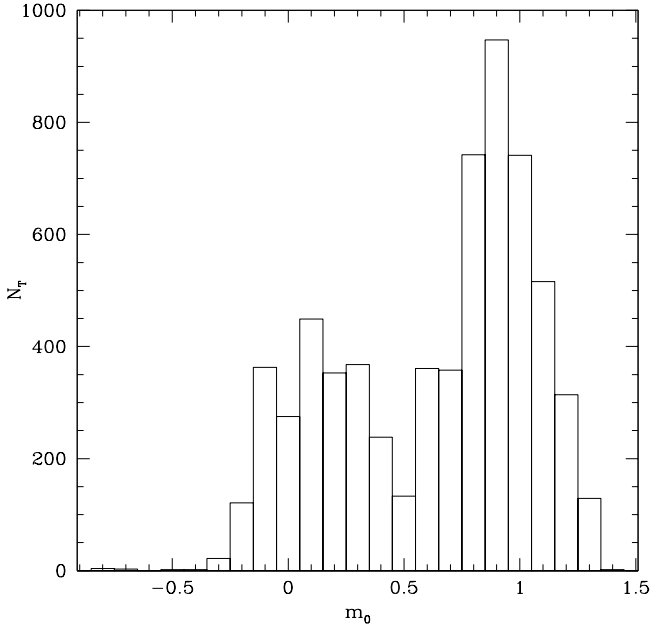}}
\caption{\label{fig:bimodal} Histogram of the number of tracks, $N_T$, that 
reproduce OGLE V-band data as
 well as the chromatic data set, as a function of the magnification 
ratio between images A and B, $m_0$.}
\end{figure}

With the aim of evaluating the likelihood of reproducing the microlensing and chromatic microlensing magnifications we
observed, a ``chromatic map'' for image A  was built (Figure \ref{fig:idl}). To each point in the map with
coordinates ($\Delta M_{V}$, $\Delta M_{V-I}$), corresponds a value of the probability of occurrence. Here 
$\Delta M_{V}\equiv -2.5 \log \mu_V$ represents the microlensing magnification in V-band, and $\Delta M_{V-I}$ is the difference in magnitude $m_{V}-m_{I}$ 
due to the disk size-wavelength dependence (i.e. chromatic microlensing). To compute those magnitude differences, since 
we are considering an extended object, the pattern was convolved with the brightness distribution 
of the source and, as before, we assumed a Gaussian profile. At $\lambda = 5430$ \AA 
\ (V-band) a Gaussian width of $r_{s}=0.03 \ r_{E}$ ($\sim 1.5$ light days) was adopted which corresponds to 
the average value found by Kochanek (2004). At $\lambda = 8130$ \AA \ (I-band), assuming that the accretion disk
has a size-wavelength dependence that goes as $r_{s}\varpropto \lambda^{4/3}$ (thin-disk model), the pattern was convolved
with a Gaussian width of $r_{s}=0.05 \ r_{E}$ ($\sim 3$ light days).

Once the quantities $\Delta M_{V}$ and $\Delta M_{V-I}$ were evaluated at each pixel 
in the V-band convolved pattern
and in the V-I difference pattern respectively, the probability of occurrence was calculated, and its distribution is
represented in Figure \ref{fig:idl}.

\begin{figure*} 
\centerline{\includegraphics[width=4.5in]{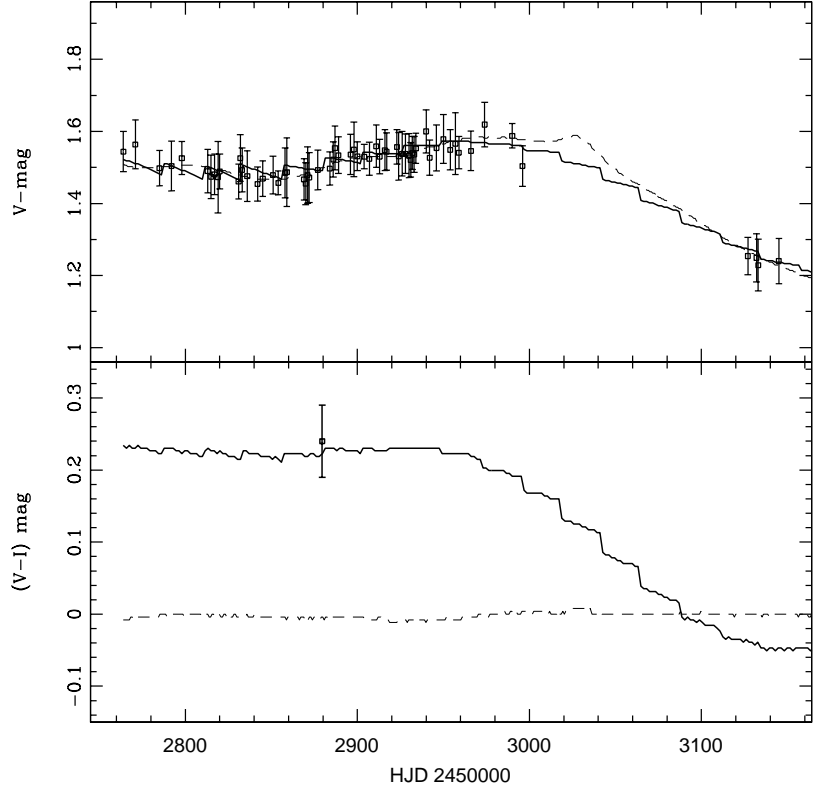}}
\caption{\label{fig:chrom} In the top panel, the best OGLE light curve fitting 
(dashed line; $\chi^2=10$) is shown together with the fitting that also reproduces 
the chromaticity ($\chi^2=13$; solid line). In the lower panel the 
chromatic effects produced along each trajectory are compared, where the
filled square represents the observed chromaticity.}
\end{figure*}

\begin{figure*}
\begin{minipage}{8.cm}
\hfill
\includegraphics[width=5.cm]{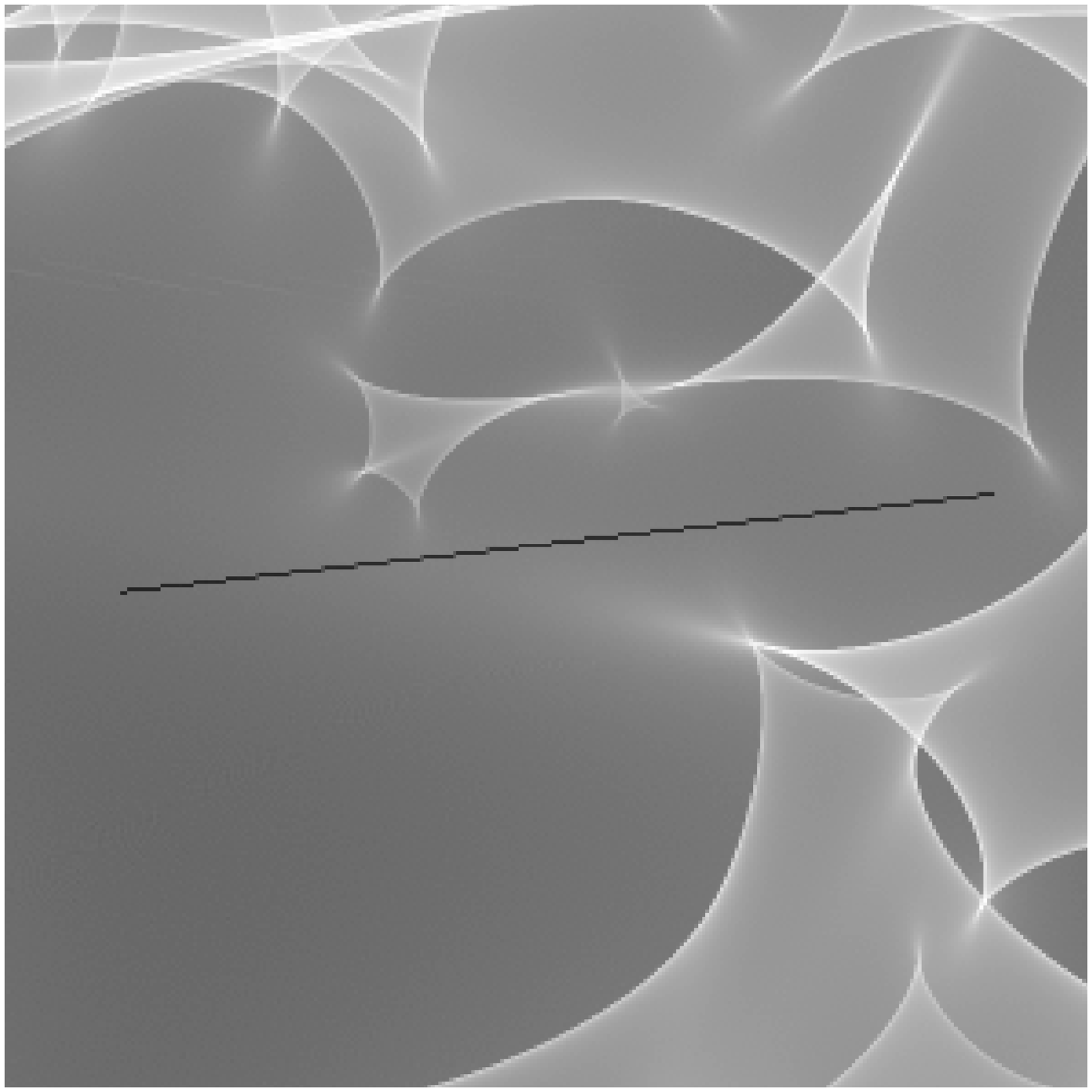}
\end {minipage}
\begin{minipage}{5.cm}
\includegraphics[width=5.cm]{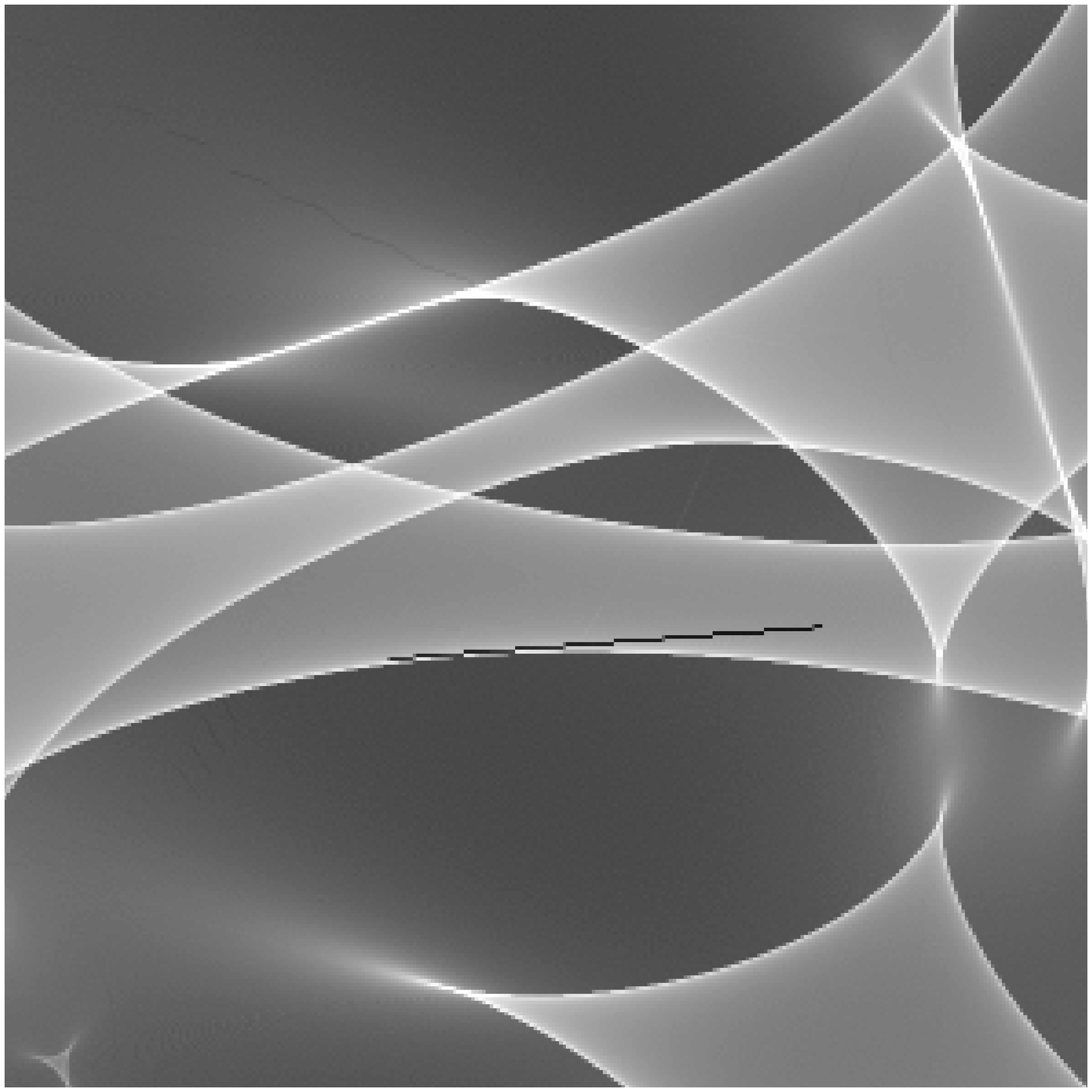}
\end{minipage}
\hfill
\caption{\label{fig:pattern} {\it Left}: Trajectory that best fits the OGLE data, but which does not reproduce the 
observed chromaticity. {\it Right}: Trajectory that best reproduces both data sets. Both sections of the magnification pattern 
have a physical size of $2.9 \ r_E \times 2.9 \  r_E$.}
\end{figure*}

As far as the baseline for no microlensing ({\it i.e.}, the magnification ratio 
between the images in the absence of microlensing) is unknown; the true value of $\Delta M_{V}$  
is unknown. However, our observations in the filters Str-u and Str-v which are affected by the 
emission lines
would allow us, in principle, to estimate separately the true magnification ratio
between the images, and the microlensing. This is because the observed flux
for the image $i$ is given by $F_{i}= \mu^{macro}_i(\mu^{micro}_{i}F_{c} + F_{l})$, where the 
subindex $c$ refers to the 
continuum and $l$ to the emission lines, {\it i.e.}, the microlensing is affecting the continuum but not
the emission lines. In practice this calculation is complicated to
do because a model for the continuum and the emission lines have to be assumed, and also because 
small variations in the measured magnitudes introduce large 
uncertainties in both factors ($\mu^{macro}_i$ and $\mu^{micro}_i$). For instance, 
these large uncertainties
do not allow the determination of the microlensing in image C. For image A, 
using our measurement in the Str-u filter and modeling the Ly${\alpha}$ emission 
line and its nearby continuum according to
the SDSS quasar composite spectrum (Vanden Berk et al. 2001),
we estimate a flux ratio $F_B/F_A\sim F_D/F_A\sim 0.6\pm 0.2$ and a microlensing 
magnification of $\mu_V \sim -1.2\pm 0.4$ mag.
These flux ratios are in agreement with those found by Wayth et al. (2005). 
Moreover a similar macroamplification for the four images 
would be a reasonable scenario for a macrolens model describing the large symmetry 
observed in the four image positions, 
but it would imply that the anomalous brightness of the image A is produced
by a large microamplification background.

We found that given a microlensing magnification of $\mu_V = -1.2 \pm 0.4 $ mag for image A, 
the probability of observing a chromatic microlensing magnification $m^V_A - m^I_A =  -0.24 \pm 0.05$ mag is of $\sim 4 \%$.
Even though this value is not very high, the probability of finding a solution is now different from zero. 
That is, the modeling of microlensing  
chromaticity from microlensing magnification maps leads to a  
phenomenology (richer than that associated with the caustic crossing  
approximation) that includes the observed data. 
The square in the ``chromatic map''(Figure  \ref{fig:idl}) corresponds to these values. The probability increases if, instead of using the $r_s \varpropto \lambda^{4/3}$ 
law, greater values of $r^I_s/r^V_s$ are considered, reaching a maximum 
value of $\simeq 27 \%$ for $r^I_s/r^V_s\simeq 3.9$.

The microamplification estimation of $\mu_V = -1.2 \pm 0.4$ mag could be affected by large uncertainties;
for instance a real source 
spectrum with deviations from the Gaussian approximation of the emission line 
profile (Vanden Berk al. 2001), or dust in the lens galaxy, 
would lead to different flux ratios between the images and microamplification values, and therefore it must be only considered as a tentative estimation.
A spectroscopic monitoring of the system would be needed to clarify the true magnification ratios and microlensing for the images.

To complete our study we explore the possibilities of jointly reproducing the $m_B-m_A$ V-band OGLE data  
and our observations. In the first place, we fit the OGLE data using the 
V-band convolved pattern, where $2\times 10^{6}$ 
random trajectories were traced, and compared with the OGLE difference $m_B-m_A$  
through a $\chi^{2}$ statistics. The source plane velocity, 
and the macromagnified image difference  were varied in the ranges 0.0015$ \ r_E$ HJD$^{-1} < v< 0.0065 \ r_E$ HJD$^{-1}$, 
and -1.08 mag $< m_0< $ 2.59 mag respectively. The wide range of the covered source plane 
velocities includes already 
existing limits (Kochanek 2004), and the bounds set for the macromagnified image 
difference were chosen according 
to covering a large range of microlensing magnifications (from $-2.48$ mag to $1.19$ mag ). 
From a total of  $7.16\times 10^{10}$ 
simulated trajectories we have found $2711,493$ tracks with $\chi^{2}/N_{f}< 1$ 
(where $N_f =57$) which satisfactorily fit the OGLE data.

The next step is to determine in how many of those tracks the observed chromaticity 
($m^V_{A}-m^I_{A}=-0.245\pm 0.05$) 
is reproduced. We found that 6443 of the tracks which reproduce OGLE data also reproduce, 
at the $1 - \sigma$ level, our
chromatic observation. The probability distribution of $m_0$ considering those trajectories 
is bimodal (Figure \ref{fig:bimodal}), with 
peaks located at $m_0 \sim 0.1$ and $m_0 \sim 0.9$, which corresponds to microlensing 
magnification values of
$\mu_V \sim -1.4$ and $\mu_V \sim -0.6$ respectively. Solutions are found in all the velocity 
range selected above with the
 number of tracks increasing when the velocity decreases. The best fit is obtained with 
$v_0=0.003\ r_E$ HJD$^{-1}$ and $m_0=0.87$ mag 
which implies $\mu_V=-0.62$ mag. In Figure  \ref{fig:chrom} (top panel) the OGLE 
 fit using these parameters is shown (solid line; $\chi^2=13$), and compared with the best 
OGLE data fitting which does not reproduce 
the chromaticity (dashed line; $\chi^2=10$). 
The chromatic behavior along each of these tracks is compared in Figure  \ref{fig:chrom} 
(bottom panel). As shown in Figure  \ref{fig:pattern}, 
both trajectories are located in a complex caustic configuration.

\begin{figure} 
\centerline{\includegraphics[width=3.5in]{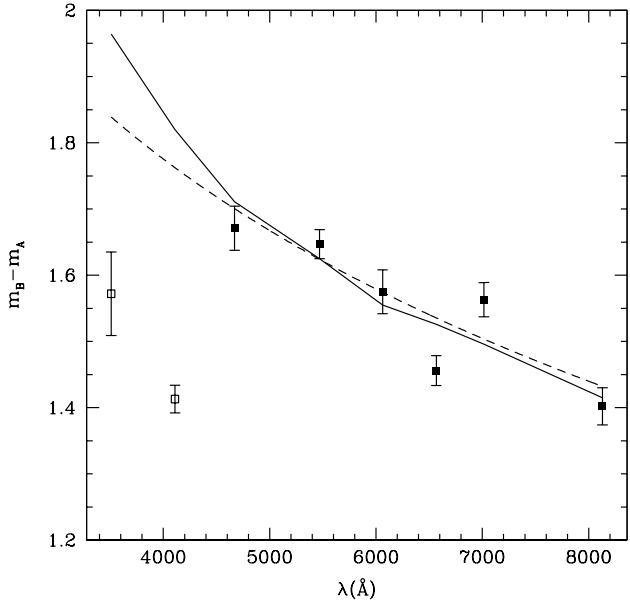}}
\caption{\label{fig:convolved} Magnitude difference between the images A and B of Q~2237+030 as a function of wavelength. 
The filled squares represent the observed values that are not affected by emission lines, while the
empty squares are associated with the two filters contaminated by emission lines. The solid line connects the 
values obtained for the date of our observation considering the track that best fits both 
data sets. These values were calculated  by convolving image A pattern with different 
source sizes corresponding to each of the filters we used, and we assumed an 
accretion disk size-wavelength scaling that goes as 
$r_{s}\sim \lambda^{4/3}$. For comparison, we plot (dashed line) a local fit to the data modeling the chromatic 
microamplification as a power law ($\mu(\lambda)\sim \lambda^{-0.44}$).}
\end{figure}

Therefore we have found models that describe both data sets under a thin disk model 
profile. Finally to compensate for our lack 
of a temporal sampling we use the good resolution in wavelength of our observations 
to check the reliability of the 
thin-disk model. Considering that the size-wavelength scaling goes as 
$r_{s}\varpropto \lambda^{4/3}$, we have convolved 
the image A magnification pattern with Gaussian profiles of different 
$r_s$ corresponding to each of our filters. 
For the track that best fits the data sets, the chromatic microlensing magnification was computed 
for the day of our observation in each filter. The values predicted by 
this model are in perfect agreement with the
observed ones that correspond to filters not affected by emission lines 
(see Figure  \ref{fig:convolved}, solid line).  This does not prove that
the $r_{s}\varpropto \lambda^{4/3}$ law is the best model fitting the
data however, it is remarkable, that such a model is fully consistent
with our observations.

Needless to say, other accretion disk size-wavelength relationships will also lead to good solutions of both data
sets. In fact, bigger  $r^I_s/r^V_s$ ratios produce a larger number of tracks that describe the
observed chromaticity and microlensing magnifications correctly. Therefore we show that our observations are 
compatible with a thin-disk model, although laws with larger $r^I_s/r^V_s$ ratios would have higher 
probability because they require small microlensing backgrounds. In any case a single chromatic event does not 
allow us to constrain enough the variation
of the source size with wavelength, what can only be done with several chromatic microlensing detections at different epochs, 
and performing a global fit to the data. For this reason we are developing an ongoing project where Q~2237+0305 is monitored with the same filter set
used in this work.

\section{Conclusions}

Using excellent quality ground-based data (narrowband photometry  
taken with the NOT at a single epoch) we have detected microlensing in  
the continuum but not in the emission lines in two images of  
Q~2237+0305 (A and C). This detection confirms that in luminous  
quasars such as Q~2237+0305 the BLR is large enough to escape global  
microlensing magnification.  The data unambiguously indicate that the  
microlensing undergoing by image A is chromatic.  This is an 
interesting detection, since this phenomenology is a fundamental tool to  
understand the structure of quasar accretion disks.

Using our data set together with the OGLE data, we have  
explored different physical scenarios in which both data sets can be  
fitted. The observations cannot be reproduced by the simple  
phenomenology described under the caustic crossing approximation.  
Modeling microlensing from magnification patterns and considering a  
Gaussian brightness profile for the accretion disk we found solutions  
compatible with a simple thin-disk model. The best solution matches  
remarkably well the variation of  microlensing magnification with  
wavelength according to the $r_{s}\varpropto \lambda^{4/3}$ law.

However, the low probability of the solutions leads us to think that the  
considered microlensing/quasar models could not be enough complete or  
realistic to describe the phenomenology of chromaticity. In fact we  
have seen that the probability of the solutions increases for laws  
with larger $r^I_s/r^V_s$ ratios than the  $4/3$ law.   
The single chromatic event detected in Q~2237+0305 looks insufficient to  
explore more complex microlensing scenarios. Additional chromatic  
detections (expected from our ongoing narrowband monitoring of  
Q~2237+0305) are needed to obtain an accurate constraint in the size-wavelength 
relation of the accretion disk for this lens system.


\acknowledgments
We thank E. E.  Falco and R. Gil-Merino for discussions on microlensing magnification models and quasar structure.
We also would like to thank the anonymous referee for very valuable comments on the manuscript.
This research was supported by the European Community's Sixth Framework Marie Curie 
Research Training Network Programme, Contract No. MRTN-CT-2004-505183 ``ANGLES'', and by the Spanish Ministerio 
de Educaci\'{o}n y Ciencias (grants AYA2004-08243-C03-01/03  and AYA2007-67342-C03-01/03).




\end{document}

%% file: tab1.tex
\setlength{\tabcolsep}{0.015in}%
\begin{deluxetable}{lccc}
\tabletypesize{\scriptsize}
\tablecaption{\label {log_NOT}Log of ALSFOC Observations}
\tablewidth{3.5in}
\tablehead{\colhead{Target} & \colhead{Observation Date} & \colhead{Filter} & \colhead{Exposure (s)}}
\startdata
Q~2237+0305 & 2003 Aug 26 & Str-u ($\lambda=3510$ \AA)& $4\times600 $\\
Q~2237+0305 & 2003 Aug 26 & Str-v ($\lambda=4110$ \AA) & $3\times600$ \\
Q~2237+0305 & 2003 Aug 26 & Str-b ($\lambda=4670$ \AA) & $3\times600$ \\
Q~2237+0305 & 2003 Aug 26 & Str-y ($\lambda=5470$ \AA) & $3 \times600$\\
Q~2237+0305 & 2003 Aug 26 & Iac\#28 ($\lambda=6062$ \AA) & $3\times600$ \\
Q~2237+0305 & 2003 Aug 26 & H$\alpha$ ($\lambda=6567$ \AA) & $3\times600$ \\
Q~2237+0305 & 2003 Aug 26 & Iac\#29 ($\lambda=7015$ \AA) & $3\times600$\\
Q~2237+0305 & 2003 Aug 26 & I-band ($\lambda=8130$ \AA) &  $4\times100$\\
Q~2237+0305 & 2003 Aug 28 & Str-u ($\lambda=3510$ \AA)& $3\times600$\\
            &             &                           & $2\times900$\\
Q~2237+0305 & 2003 Aug 28 & Str-v ($\lambda=4110$ \AA) & $3\times600$ \\
Q~2237+0305 & 2003 Aug 28 & Str-b ($\lambda=4670$ \AA) & $3\times420$ \\
Q~2237+0305 & 2003 Aug 28 & Str-y ($\lambda=5470$ \AA) & $3\times420$ \\
Q~2237+0305 & 2003 Aug 28 & Iac\#28 ($\lambda=6062$ \AA) & $3\times420$ \\
Q~2237+0305 & 2003 Aug 28 & H$\alpha$ ($\lambda=6567$ \AA) & $3\times420$ \\
Q~2237+0305 & 2003 Aug 28 & Iac\#29 ($\lambda=7015$ \AA) & $3\times420$ \\
Q~2237+0305 & 2003 Aug 28 & I-band ($\lambda=8130$ \AA) & $3\times100$ \\
\enddata
\end{deluxetable}

%% file: tab2.tex
\setlength{\tabcolsep}{0.1in}%
\begin{deluxetable}{lccccccc}
\tabletypesize{\scriptsize}
\tablecaption{Q~2237+0305 PHOTOMETRY}
\tablewidth{3.5in}
\tablehead{\colhead{Filter} & \colhead{$m_B-m_A$} & 
\colhead{$m_C-m_A$} & \colhead{$m_D-m_A$}}
\startdata
Str-u ($\lambda=3510$ \AA)& 1.57$\pm$0.06 & 1.57$\pm$0.06 & 1.52$\pm$0.06  \\
Str-v ($\lambda=4110$ \AA) & 1.41$\pm$0.02 & 1.45$\pm$0.04 & 1.39$\pm$0.02 \\
Str-b ($\lambda=4670$ \AA) & 1.67$\pm$0.03 & 1.47$\pm$0.04 & 1.59$\pm$0.04  \\
Str-y ($\lambda=5470$ \AA) & 1.65$\pm$0.02 & 1.45$\pm$0.06 & 1.58$\pm$0.03  \\
Iac\#28 ($\lambda=6062$ \AA) & 1.58$\pm$0.03 & 1.39$\pm$0.04 & 1.52$\pm$0.04 \\
H$\alpha$ ($\lambda=6567$ \AA) & 1.46$\pm$0.02 & 1.32$\pm$0.03 & 1.46$\pm$0.05 \\
Iac\#29 ($\lambda=7015$ \AA) & 1.56$\pm$0.03 & 1.38$\pm$0.02 & 1.52$\pm$0.02\\
I-band ($\lambda=8130$ \AA) & 1.40$\pm$0.03 & 1.22$\pm$0.02 & 1.38$\pm$0.04  \\
\enddata
\end{deluxetable}